\documentclass[prl,aps,twocolumn,preprintnumbers,showpacs,showkeys,superscriptaddress]{revtex4}

\usepackage{epsfig,amssymb,amsfonts,verbatim}

\def\bfl{\begin{flushleft}}
\def\efl{\end{flushleft}}
\def\bfr{\begin{flushright}}
\def\efr{\end{flushright}}
\def\bc{\begin{center}}
\def\ec{\end{center}}
\def\be{\begin{equation}}
\def\ee{\end{equation}}
\def\ba{\begin{eqnarray}}
\def\ea{\end{eqnarray}}
\def\baa#1{\begin{array}{#1}}
\def\eaa{\end{array}}
\def\bw{\begin{widetext}}
\def\ew{\end{widetext}}
\def\nn{\nonumber }

\def\text#1{\mbox{#1}}

\def\square{{\hbox{$\sqcup$}\llap{\hbox{$\sqcap$}}}}

\begin{document}


\title{Hall resistance and Lorentz ratio models in YBa$_2$Cu$_3$O$_7$}

\author{Andrew Das Arulsamy}

\affiliation{ Condensed Matter Group, Division of Exotic Matter,
No. 22, Jalan Melur 14, Taman Melur, 68000 Ampang, Selangor DE,
Malaysia}

\date{\today}



\begin{abstract}
Two-dimensional models of heat capacity, conductivity ($\kappa$),
Hall resistance ($R_H$) and the Lorentz ratio ($\mathcal{L}$) have
been derived using the ionization energy ($E_I$) based Fermi-Dirac
statistics (iFDS) for overdoped Cuprates. These models reproduce
the experimentally measured $\rho(T)$ and $R_H(T)$. The variation
of $\mathcal{L}$ is weakly $T$-dependent due to the experimental
$\kappa(T)$. The $e$-$ph$ coupling in the electrical resistivity
has the polaronic effect that complies with iFDS, rather than the
$e$-$ph$ scattering, which satisfies the Bloch-Gr\"{u}neisen
formula. These models are purely Fermi liquid and are not
associated with any microscopic theories of high-$T_c$
superconductors.
\end{abstract}



\maketitle



\subsection{1. Introduction}

The electrical properties of high-$T_c$ Cuprate superconductors
(HTSC) are intrinsically enigmatic in both experimental and
theoretical framework compared to other oxide compounds, including
manganites. Partly due to its huge potential in a wide variety of
applications, intense focus is given on the nature of conductivity
of these materials to shed some light on the puzzling
temperature-dependence issues in heat capacity ($\mathcal{C}$),
heat conductivity, resistivity, Hall resistance and Lorentz ratio.
The conflict in term of $T$-dependency between $\rho$ and $R_H$ is
also one of the unsolved mystery in HTSC. Even though this paper
does not solve it microscopically, but it points out that the
Fermi liquid with strong correlations is not downright incorrect,
at least for over- and optimally doped HTSC. Apart from HTSC, the
applications of ionization energy ($E_I$) based Fermi-Dirac
statistics (iFDS) on ferromagnets, diluted ferromagnetic
semiconductors and doped ferroelectrics have been derived and
discussed
analytically~\cite{andrew4,andrew5,andrew6,andrew7,andrew8,andrew9,andrew10}.
The iFDS in HTSC have been successfully
exploited~\cite{andrew4,andrew5,andrew6} via the experimental data
reported in the
Refs.~\cite{abd11,naqib12,batista13,meen14,khosroabadi15,isawa16,shi17,kandyel18,karimoto19,sulkowski20,zheng21,hagen22}.
Recently, the said puzzling $T$ dependencies as well as the spin
gap phenomenon have been tackled with the coupling of iFDS and
charge spin separation~\cite{andrew1,andrew2,andrew3}.
Unfortunately, the pure charge-spin separation, is believed to
have serious
shortcomings~\cite{varma30,varma31,alex32,alex33,lanzara34}. In
addition, there are also numerous experimental reports with
controversial interpretations surrounding the spin Pseudogap
phenomenon~\cite{andrew3,krasnov23,takenaka24,anderson25,anderson26,anderson27,maekawa28,kleefisch29}.
As such, by ignoring the spin Pseudogap phase, the
thermo-magneto-electronic properties of YBa$_2$Cu$_3$O$_7$ will be
discussed based on the iFDS by heavily relying on the basic
transport experiments such as the resistivity, Hall resistance,
heat capacity and heat conductivity. It is interesting to note
that these purely Fermi-liquid models are able to reproduce the
related experimental data reasonably well even if they are only
for over- and optimally doped HTSC. The polaronic effect that
arises as a result of iFDS is solely due to heavier effective mass
effect, which \textit{could} indicate the existence of polarons.
But this indication is just an extrapolated assumption since heavy
electrons do not necessarily form polarons.

\subsection{\textbf{2. Theoretical details}}

The free-particle Hamiltonian of mass $m$ moving in 3-dimensions
is given by

\begin {eqnarray}
\hat{H} = \frac{\hat{p}^2}{2m} = -\frac{\hbar^2}{2m}\nabla^2.
\label{eq:1001}
\end {eqnarray}

Here, we have make use of the linear momentum operator,
$\hat{\textbf{p}} = -i\hbar\nabla$. Subsequently, one can write
the time-independent Schr\"{o}dinger equation for the same
particle, however in an unknown potential, $V(x)$ as

\begin {eqnarray}
-\frac{\hbar^2}{2m}\nabla^2\varphi &&= (E + V(x))\varphi \nn
\\&& = (E_0 \pm \xi)\varphi . \label{eq:1002}
\end {eqnarray}

In the second line of Eq.~(\ref{eq:1002}), one can notice that the
influence of the potential energy on the total energy of that
particular particle has been conveniently parameterized as $\xi$.
This energy function, $\xi$ will be characterized later in such a
way that one can replace $E + V(x)$ with $E_0 + \xi$ in which,
$E_0$ = $E$ at $T$ = 0. Add to that, from Eq.~(\ref{eq:1002}), it
is obvious that the magnitude of $\xi$ is given by $\pm\xi = E -
E_0 + V(x)$. Physically, it implies the energy needed to overcome
the potential energy as well as the bound state. Literally, this
is exactly what we need to know in any condensed matter, i.e.,
this magnitude is the one that actually or reasonably defines the
properties of the quasiparticles. Subsequently, we obtain

\begin {eqnarray}
\nabla^2\varphi = -\frac{2m}{\hbar^2}[E_0 \pm \xi]\varphi.
\label{eq:1003}
\end {eqnarray}

\begin {eqnarray}
\frac{\hbar^2k^2}{2m} = E_0 \pm \xi = \frac{\hbar^2}{2m}[k_0^2 \pm
k_{\xi}^2]. \label{eq:1004}
\end {eqnarray}

$k^2$ = $(2m/\hbar^2)[E_0 \pm \xi]$. $E$ and $E_0$ in a given
system range from $+\infty$ to 0 for electrons and 0 to $-\infty$
for holes that eventually explains the $\pm$ sign in $\xi$. Now,
Eq.~(\ref{eq:1002}) can be solved to give

\begin {eqnarray}
&&\varphi = C_N \times \nn \\&& \exp[i(k_{0,x} \pm k_{\xi,x})x +
i(k_{0,y} \pm k_{\xi,y})y + i(k_{0,z} \pm k_{\xi,z})z]. \nn \\&&
\varphi_{\textbf{k}(0,\xi)} =
C_Ne^{i\textbf{k}(0,\xi)\cdot\textbf{r}}. \label{eq:1005}
\end {eqnarray}

$k^2 = (k^2_{0,x} \pm k^2_{\xi,x}) + (k^2_{0,y} \pm k^2_{\xi,y}) +
(k^2_{0,z} \pm k^2_{\xi,z})$. By employing the orthonormality and
Plancherel's theorem, one can find the normalization constant,
$C_N$ by comparing Eqs.~(\ref{eq:1006}) and~(\ref{eq:1007}) as
shown below.

\begin {eqnarray}
&&\langle\varphi_{\textbf{k}_0}\mid\varphi_{\textbf{k}_0 \pm
\textbf{k}_{\xi}}\rangle =
\int\int\int\varphi_{\textbf{k}_0}*\varphi_{\textbf{k}_0\pm\textbf{k}_{\xi}}
dx dy dz \nn
\\&& = C_N^2 \int\int\int e^{i[\textbf{k}_0 -(\textbf{k}_0\pm \textbf{k}_{\xi})]\cdot
\textbf{r}}dx dy dz \nn
\\&& = \delta(\mp \textbf{k}_{\xi}).
\label{eq:1006}
\end {eqnarray}

\begin {eqnarray}
\frac{1}{(2\pi)^3}\int\int\int e^{i\textbf{k}_0 \cdot (\textbf{r}
- \textbf{r}')} dk_x dk_y dk_z = \delta(\textbf{r} - \textbf{r}').
\label{eq:1007}
\end {eqnarray}

Hence, $C_N = 1/(2\pi)^{3/2}$. finally, the normalized wave
function, which corresponds to Eq.~(\ref{eq:1002}) is

\begin {eqnarray}
\varphi_{\textbf{k}(0,\xi)} =
\frac{1}{(2\pi)^{3/2}}e^{i[\textbf{k}(0,\xi)]\cdot\textbf{r}}.
\label{eq:1008}
\end {eqnarray}

In a physical sense as stated earlier, $\xi = E - E_0 + V(x)$, is
in an identical scale with the energy needed to free an electron
from an atom in a given crystal. As such, we apply the concept of
ionization energy where, $\xi$ = $E_I^{real}$ = $E_I + V(x)$, to
justify that an electron to occupy a higher state $N$ from initial
state $M$ is more probable than from initial state $L$ if
condition $E_I(M)$ $<$ $E_I(L)$ at certain $T$ is satisfied. As
for a hole to occupy a lower state $M$ from initial state $N$ is
more probable than to occupy state $L$ if the same condition above
is satisfied. It is well known that the exact values of $E_I$ are
known for an isolated atom. In this case (for an isolated atom),
$E_I$ can be evaluated with

\begin {eqnarray}
&E_I& = \sum_i^z \frac{E_{Ii}}{z}.\label{eq:1009}
\end {eqnarray}

However, substituting the same atom in a crystal gives rise to the
influence of $V(x)$ and in reality, $E_I^{real}$ cannot be
evaluated from Eq.~(\ref{eq:1009}). Nevertheless, the $E_I^{real}$
of an atom or ion in a crystal is proportional to the isolated
atom and/or ion's $E_I$ as written below.

\begin {eqnarray}
&E_I^{real}& = \alpha\sum_i^z \frac{E_{Ii}}{z} \nn \\&& = \alpha
E_I .\label{eq:1010}
\end {eqnarray}

It is this property that enables one to predict the variation of
electronic properties of superconductors with substitution
reasonably well. The constant of proportionality, $\alpha$ is a
function of averaged $V(x)$ and varies with different background
atoms. For example, in YBa$_{1-x}$Ca$_x$Cu$_2$O$_7$ system,
YBa$_{1-x}$-Cu$_2$O$_7$ defines the background atoms or ions.
Therefore, one needs to employ the experimental data to determine
the magnitude of $\xi$ = $E_I^{real}$ = $E_I + V(x)$.

Recall that Eq.~(\ref{eq:1004}) simply implies that the
one-particle energies $E_1$, $E_2$, ..., $E_m$ for the
corresponding one-particle quantum states $q_1$, $q_2$, ..., $q_m$
can be rewritten as ($E_0$ $\pm$ $E_I)_1$, ($E_0$ $\pm$ $E_I)_2$,
..., ($E_0$ $\pm$ $E_I)_m$. It is also important to note that
$E_0$ + $E_I$ = $E_{electrons}$ and $E_0$ $-$ $E_I$ = $E_{holes}$.
As such, for $n$ particles, the total number of particles and its
energies are conserved and the conditions to fulfill those
conservations are given by

\begin {eqnarray}
&&\sum_i^{\infty} n_i = n,~~~~~ \sum_i^{\infty} dn_i = 0.
\label{eq:1011}
\end {eqnarray}

\begin {eqnarray}
&&\sum_i^{\infty} (E_0 \pm E_I)_i n_i = E,~~~~~ \sum_i^{\infty}
(E_0 \pm E_I)_i dn_i = 0.\nn \\&&\label{eq:1012}
\end {eqnarray}

Subsequently, the Fermi-Dirac statistics based on ionization
energy can be derived as

\begin {eqnarray}
\frac {n_i}{q_i} = \frac{1}{\exp [\mu + \lambda (E_0 \pm E_I)_i] +
1}.\label{eq:1013}
\end {eqnarray}

By utilizing Eq.~(\ref{eq:1013}) and taking $\exp[\mu + \lambda(E
\pm E_I)]$ $\gg$ 1, one can arrive at the probability function for
electrons in an explicit form as

\begin{eqnarray}
f_e = \exp \left[-\mu-\lambda\left(\frac{\hbar^2{\bf
k}_0^2}{2m}+E_I\right) \right], \label{eq:1015}
\end{eqnarray}

Similarly, the probability function for the holes is given by

\begin{eqnarray}
f_h = \exp\left[\mu + \lambda\left(\frac{\hbar^2{\bf
k}_0^2}{2m}-E_I\right) \right]. \label{eq:1016}
\end{eqnarray}

The parameters $\mu$ and $\lambda$ are the Lagrange multipliers.
$\hbar$ $=$ $h/2\pi$, $h$ $=$ Planck constant and $m$ is the
charge carriers' mass. In the standard FDS, Eqs.~(\ref{eq:1015})
and~(\ref{eq:1016}) are simply given by, $f_e({\bf k})$ $=$
$\exp[-\mu-\lambda(\hbar^2{\bf k}^2/2m)]$ and $f_h({\bf k})$ $=$
$\exp[\mu+\lambda(\hbar^2{\bf k}^2/2m)]$ respectively.

Subsequently Eq.~(\ref{eq:1011}) can be rewritten by employing the
2D density of states' (DOS) derivative, $dn$ $=$ $L^2{\bf
k}_0d{\bf k}_0/2\pi$, Eqs.~(\ref{eq:1015}) and~(\ref{eq:1016}),
that eventually give

\begin{eqnarray}
n & = &\frac {L^2}{2\pi}e^{-\mu - \lambda E_I}
\int\limits_0^\infty {\bf k}_0 \exp\left(-\lambda
\frac{\hbar^2{\bf k}_0^2}{2m}\right) d{\bf k}_0,\label{eq:1017}
\end{eqnarray}

\begin{eqnarray}
p & = & \frac {L^2}{2\pi}e^{\mu - \lambda E_I}
\int\limits_{-\infty}^0 {\bf k}_0 \exp\left(\lambda \frac
{\hbar^2{\bf k}_0^2}{2m}\right) d{\bf k}_0.\label{eq:1018}
\end{eqnarray}

Note here that $E_0$ is substituted with $\hbar^2{\bf
k}_0^2$/$2m$. $n$ and $p$ are the respective concentrations of
electrons and holes. $L^2$ denotes area in {\bf k}-space. The
respective solutions of Eqs.~(\ref{eq:1017}) and~(\ref{eq:1018})
are given below

\begin{eqnarray}
e^{\mu + \lambda E_I} & = &
\frac{mL^2}{2n\pi\lambda\hbar^2},\label{eq:1019}
\end{eqnarray}

\begin{eqnarray}
e^{\mu - \lambda E_I} & = &
\frac{2p\pi\lambda\hbar^2}{mL^2},\label{eq:1020}
\end{eqnarray}

Equations~(\ref{eq:1019}) and~(\ref{eq:1020}) respectively imply
that

\begin{eqnarray}
\mu_e(iFDS) = \mu_e + \lambda E_I, \label{eq:1021}
\end{eqnarray}

\begin{eqnarray}
\mu_h(iFDS) = \mu_h - \lambda E_I. \label{eq:1022}
\end{eqnarray}

On the other hand, using Eq.~(\ref{eq:1012}), one can obtain

\begin{eqnarray}
E && = \frac {L^2\hbar^2}{4m\pi}  e^{-\mu -\lambda E_I}
\int\limits_0^\infty {\bf k}_0^3
\exp\left(\frac{-\lambda\hbar^2{\bf k}_0^2}{2m}\right)d{\bf k}_0
\nn
\\&& = \frac{m}{2\pi}\bigg(\frac{L}{\lambda\hbar}\bigg)^2 e^{-\mu
-\lambda E_I}. \label{eq:1023}
\end{eqnarray}

Equation~(\ref{eq:1023}), after appropriate substitution with
Eq.~(\ref{eq:1019}) is compared with the energy of a 2D ideal gas,
$E = nk_BT$. Quantitative comparison will eventually leads to
$\lambda_{iFDS}$ = $\lambda_{FDS}$ = 1/$k_BT$ where $k_B$ is the
Boltzmann constant. The distribution function for electrons and
holes can be written explicitly by first denoting $\mu$ $=$ $-E_F$
(Fermi level), $\lambda$ $=$ 1/$k_BT$ and substituting these into
Eqs.~(\ref{eq:1015}) and~(\ref{eq:1016}) will lead one to write

\begin{eqnarray}
f_e(E,E_I) &=& \exp\left[\frac{E_F - E_I - E}{k_BT}\right].
\label{eq:1024}
\end{eqnarray}

\begin{eqnarray}
f_h(E,E_I) &=& \exp\left[\frac{E - E_I - E_F}{k_BT}\right].
\label{eq:1025}
\end{eqnarray}

Note that Eqs.~(\ref{eq:1021}) and~(\ref{eq:1022}) simply imply
that $\mu_e(iFDS)$ $=$ $\mu(T=0)$ + $\lambda E_I$ and
$\mu_h(iFDS)$ $=$ $\mu(T=0)$ $-$ $\lambda E_I$. In fact,
$\mu(FDS)$ need to be varied accordingly with doping, on the other
hand, iFDS captures the same variation due to doping with $\lambda
E_I$ in which, $\mu(T=0)$ is fixed to be a constant (independent
of $T$ and doping). Furthermore, using
Eqs.~(\ref{eq:1004}),~(\ref{eq:1012}) and~(\ref{eq:1023}), one can
obtain

\begin{eqnarray}
E_{iFDS} && = \frac {L^2\hbar^2}{4m\pi}  e^{-\mu -\lambda E_I}
\int\limits_0^\infty {\bf k}_0^3
\exp\left(\frac{-\lambda\hbar^2{\bf k}_0^2}{2m}\right)d{\bf k}_0
\nn
\\&& = \frac {L^2\hbar^2}{4m\pi}  e^{-\mu}
\int\limits_0^\infty {\bf k}_0^3
\exp\left(-\frac{\lambda\hbar^2{\bf k}_0^2}{2m} -
\frac{\lambda\hbar^2{\bf k}_\xi^2}{2m}\right)d{\bf k}_0 \nn
\\&& = \frac {L^2\hbar^2}{4m\pi}  e^{-\mu}
\int\limits_0^\infty {\bf k}^3 \exp\left(-\frac{\lambda\hbar^2{\bf
k}^2}{2m}\right)d{\bf k} \nn
\\&& = E_{FDS}. \label{eq:1026}
\end{eqnarray}

Eventually, Eq.~(\ref{eq:1026}) proves that the total energy of
$n$ particles considered in both FDS and iFDS is exactly the same.

\subsection{3. Heat capacity and its conductivity}

Electrons and phonons can be excited to a higher energy levels
satisfying the ionization energy based Fermi-Dirac ($f_{iFDS}(E)$)
and Bose-Einstein $f_{BES}(E)$ statistics respectively. Therefore,
the heat capacity can be explicitly written as

\begin{eqnarray}
\mathcal{C} &=& \frac{m^*}{\pi\hbar^2}\bigg[\int\limits^\infty_0
(E-E_F-E_I)\frac{\partial f_{iFDS}(E)}{\partial T}dE \nonumber
\\&& + \int\limits^\infty_0 (E-E_F)\frac{\partial
f_{BES}(E)}{\partial T}dE\bigg]. \label{eq:1}
\end{eqnarray}

$\hbar$ = $h/2\pi$, $h$ denotes Planck constant, while $m^*$
represents the effective mass. The respective distribution
functions for BES and iFDS (using Eq.~(\ref{eq:1004})) are given
by $f_{BES}(E) = 1/[\exp\big[(E-E_F)/k_BT\big] - 1] \approx
\exp\big[(E_F-E)/k_BT\big]$ and $f_{iFDS}(E) =
1/[\exp\big[(E-E_F+E_I)/k_BT\big] + 1] \approx
\exp\big[(E_F-E_I-E)/k_BT\big]$. These approximations are
necessary to avoid the exponential integral function, $\Xi_i(z) =
\int_{-z}^{\infty}[e^{-t}/t]dt$, which has a branch cut
discontinuity in the complex $z$ plane running from $-\infty$ to
0. Additionally, I highlight that for classical particles
satisfying the Maxwell-Boltzmann statistics (MBS), there is no
such thing as $E_I$. Consequently, one should not assume that
$\exp[\mu + \lambda(E_{initial~state} \pm E_I)]$ $\gg$ 1 should
give the MB distribution function as a classical or a
free-electron limit. One can indeed arrive at MBS by first
considering the additional constraint, $E_I$ = 0 in such cases,
where $E_{total}$ now equals to $E$ identical with the standard
FDS and MBS. Therefore, the electron's Fermi level, $E_F$ term
that contained in $f_{BES}(E)$ corresponds to the phonons'
energies above this $E_F$, in which this $E_F$ does not imply
phonons' chemical potential. In other words, phonons with energies
$<$ $E_F$ are neglected. In summary, iFDS captures the Fermi
liquid ($\xi \neq 0$) rather than the Fermi gas ($V(x) = 0$).

The total heat capacity in Eq.~(\ref{eq:1}) has been written as
$\mathcal{C}$ = $\mathcal{C}^{e}$ + $\mathcal{C}^{ph}$ as a result
of the total heat current, $\kappa =
-\sum_{\alpha}j_Q^{\alpha}(\nabla T)^{-1}$ =
$\sum_{\alpha}\mathcal{C}^{\alpha}v_F^2\tau_{\nu}/2$. $\alpha$ =
electron (\textit{e}), phonon (\textit{ph}) and $\nu$ =
\textit{e-e}, \textit{e-ph} scattering. $v_F$ denotes the Fermi
velocity and $k_B$ is the Boltzmann constant. Importantly, $E_I$
is microscopically defined as~\cite{andrew10}

\begin {eqnarray}
\epsilon(0,\textbf{k}) = 1 +
\frac{\mathcal{K}_s^2}{\textbf{k}^2}\exp\big[\lambda(E_F^0-E_I)\big].\label{eq:2}
\end {eqnarray}

$\epsilon(0,\textbf{k})$ is the static dielectric function,
$\textbf{k}$ and $\lambda$ are the wavevector and Lagrange
multiplier respectively. The $E_F^0$ denotes the Fermi level at 0
K, while the $\mathcal{K}_s$ represents the Thomas-Fermi screening
parameter. Unlike electrical resistivity in YBa$_2$Cu$_3$O$_7$,
its 2D heat conductivity is equally strongly influenced by $e$-$e$
and $e$-$ph$ interactions, hence (after taking $E_F =
\frac{1}{2}m^*v_F^2$)

\begin{eqnarray}
\kappa = \tau_{e-e}\mathcal{C}^{e}\frac{E_F}{m^*} +
\tau_{e-ph}\mathcal{C}^{ph}\frac{E_F}{m^*_{ph}}. \label{eq:3}
\end{eqnarray}

The explicit form of Eq.~(\ref{eq:3}) can be obtained after
substituting Eq.~(\ref{eq:1}) into Eq.~(\ref{eq:3}) appropriately.
The electron-electron scattering rate, $\tau_{e-e}^{-1}$ =
$\tau_e^{-1}$ = $AT^2$ while the electron-phonon scattering rate,
$\tau_{e-ph}^{-1}$ is assumed to be proportional to $T^{\alpha}$
in which $\alpha$ $>$ 2. The Fermi-level in Eqs.~(\ref{eq:1}) and
~(\ref{eq:3}) implies that the phonons considered here have the
thermal energies in the order of or higher than the electrons'
Fermi energy which eventually means that these electrons cannot
form Fermi gas. In simple words, if the thermal energies of the
phonons are less than the electrons' $E_F$, then these electrons
can act as Fermi gas and one may employ the Debye approximation.
This is another reason why Debye model works extremely well at
intermediate and low temperatures in common metals. However, the
phonons' effective mass is equal to the ions reduced mass due to
phonons interaction with free-electrons, $1/m^*_{ph} = 1/m_e +
1/m_{ion}$, which needs to be determined from other techniques, be
it theoretical or experimental. Therefore, only the electron's
effective mass is highlighted here. In other words, instead of
addressing $1/m^*_{ph}$ as the reduced mass of ions, it has been
labelled as phonons effective mass so that one can conveniently
identify it as the parameter belonging to the phonons'
contribution.

\subsection{4. Resistivity and Hall resistance}

The equations of motion (EOM) for charge carriers in $ab$-planes
under the influence of static magnetic ({\bf H}) and electric
fields ({\bf E}) can be written in an identical fashion as given
in Ref.~\cite{kittel35}, which are given by $m^*\big[d/dt +
1/\tau_e\big]v_b = e{\bf E}_b + e{\bf H}_cv_a$ and $m^*\big[d/dt +
1/\tau_e\big]v_a = e{\bf E}_a - e{\bf H}_cv_b$. The charge, $e$ is
defined as negative in the EOM above. Moreover, it is important to
realize that the existence of electrons in $ab$-planes below
$T_{crossover}$ are actually holes. The existence of holes in
$ab$-planes was discussed intensively in the
Refs.~\cite{almasan36,zhao37,das38}. The subscripts $a$, $b$ and
$c$ represent  the axes in $a$, $b$ and $c$ directions while the
subscript $ab$ represents the $ab$-planes. In a steady state of a
static {\bf H} and {\bf E}, $dv_a/dt$ = $dv_b/dt$ = 0 and $v_a$ =
0 hence one can obtain ${\bf E}_a = e{\bf H}_c{\bf
E}_b\tau_e/m^*$. The Hall resistance and current along $a$- and
$b$-axes are respectively defined as $R_H^{(a)} = {\bf
E}_a/j_b{\bf H}_c$, $j_b = {\bf E}_b/\rho$ in which,
$\tan\theta_H^{(a)} = {\bf E}_a/{\bf E}_b$. Parallel to this,
$R_H^{(a)} = \tan\theta_H^{(a)}\rho/{\bf H}_c$. $j_b$ is the
current due to holes motion along $b$-axis and $\theta_H^{(a)}$ is
the Hall angle in $ab$-planes. Furthermore, one can rewrite
$\tan\theta_H^{(a)}$ as $\tan\theta_H^{(a)} = e{\bf H}_c/m^*AT^2$,
which eventually suggests, $\cot\theta_H^{(a)}$ $\propto$ $T^2$.
$A$ is $\tau_e$ dependent constant and is independent of $T$. The
2D resistivity model, $\rho(T)$ is given
by~\cite{andrew4,andrew5,andrew6}

\begin{eqnarray}
\rho(T) &=&
A\frac{\pi\hbar^2}{k_Be^2}T\exp\bigg[\frac{E_I+E_F}{T}\bigg].
\label{eq:4}
\end{eqnarray}

Utilizing Eq.~(\ref{eq:4}), one can show that the Hall resistance
is given by

\begin{eqnarray}
R_H =
\frac{\pi\hbar^2}{m^*Tk_Be}\exp\bigg[\frac{E_I+E_F}{T}\bigg].
\label{eq:5}
\end{eqnarray}

Thus, it is clear that $R_H$ is proportional to 1/$T$ regardless
of the axes. Detailed analysis and diagnosis of Eq.~(\ref{eq:4})
with a wide variety of experimental data are well documented in
the Refs.~\cite{andrew4,andrew5,andrew6}. Optimally doped
YBa$_2$Cu$_3$O$_{7-\delta}$ single crystal (A1) obtained from
Ref.~\cite{harris} will be utilized in the following analysis.
Equation~(\ref{eq:4}) has been employed to theoretically reproduce
(indicated with a solid line in the inset of Fig.~\ref{fig:1}) the
experimental $\rho_{ab}(T)$ by varying the $T$-independent
scattering rate constant, $A$ (7.3 $\times$ 10$^{-7}$ $\Omega
\cdot$cm) whereas $E_I + E_F$ = $T_{crossover}$ ($T_{cr}$) is
taken as 0 K, since any optimally doped YBa$_2$Cu$_3$O$_7$ gives
$T_{cr}$ $\ll$ $T_c$ in which $T_{cr}$ is not observable from the
resistivity measurements. I.e., $T_{cr}$ cannot be predicted
accurately from the normal state resistivity measurements. On the
other hand, the $R_H^{(ab)}(T)$ data and the plot using
Eq.~(\ref{eq:5}) are depicted in Fig.~\ref{fig:1}. Note that $A =
A\pi\hbar^2/k_Be^2$ from Eq.~(\ref{eq:4}) and $A_H =
\pi\hbar^2/m^*k_Be$ = 347 JKCs$^2$kg$^{-1}$ from Eq.~(\ref{eq:5}).
In the latter approximation, $m^*$ = 50$m_0$, $m_0$ is the rest
mass of the electron. In order to accurately fit the experimental
$R_H^{(ab)}(T)$ data, the effective mass should be equal to
73$m_0$, which in turn gives the charge carriers density as $p$ =
8 $\times$ 10$^{22}$ cm$^{-3}$, in accordance with the
Refs.~\cite{andrew6,sulkowski20}.

\subsection{5. Lorentz ratio}

Lanzara {\it et al}.~\cite{lanzara34} have shown that the $e$-$ph$
coupling is somewhat inevitable, which has been observed via ARPES
technique. Indeed this supports the notion of polaronic effect
above $T_c$ in cuprates. One should note that the observation
$e$-$ph$ coupling does not mean that there is a $e$-$ph$
scattering since normal state $\rho(T)$ measurements thus far
failed to reveal any $e$-$ph$ scattering (strong $T$-dependence).
Actually, this is not because of $\rho(T)$'s blindness, but due to
polaronic effect represented by $E_I$, which gives rise to the
effective mass ($m^*$) of electrons instead of strong
$T$-dependence. The heavier $m^*$ implies the existence of
polaronic effect in the normal state of HTSC that also suppresses
$e$-$ph$ scattering but not the $e$-$ph$ coupling in term of
polaronic effect. Similarly, isotope effect ($^{18}$O, $^{16}$O)
in cuprates~\cite{hofer39,iyo40,keller41} also reinforces the
polaronic contribution via $e$-$ph$ coupling rather than $e$-$ph$
scattering. In fact, Hofer {\it et al}.~\cite{hofer39} claimed
that $m^*$ reduces towards the optimally doped HTSC. This scenario
is consistent with iFDS based models that predicts $T_{cr}$ also
reduces towards optimal doping. Simply put, reduced $E_I$ will
eventually lead to reduced $m^*$ and consequently the influence of
isotope doping on $m^*$ is less effective in optimally doped
regime as compared to under doped. The inappropriateness of the
$e$-$ph$ scattering in YBCO$_7$ will be discussed in detail based
on the Bloch-Gr\"{u}neisen formula shortly. From the definition,
$\mathcal{L}$ can be written as

\begin{eqnarray}
\mathcal{L} = \frac{\rho}{T}\kappa = \kappa
A\frac{\pi\hbar^2}{k_Be^2}\exp\bigg[\frac{E_I + E_F}{T}\bigg].
\label{eq:6}
\end{eqnarray}

Interestingly, Sutherland {\it et al}.~\cite{sutherland42} have
reported only a slight increase (upward deviation) in $ab$-plane's
heat conductivity with phonon contribution ($\kappa_{ab}^{100 K}
$/$\kappa_{ab}^{300 K}$ $\le$ 1.3) above critical temperature
($T_c$) for overdoped YBCO. Their results will be used to discuss
the accuracy of Eqs.~(\ref{eq:3}) and~(\ref{eq:6}) to capture the
experimental data.

\subsection{6. $e$-$ph$ scattering in resistivity}

Firstly, the Bloch-Gr\"{u}neisen (BG) formula will be revisited in
order to rule out the $e$-$ph$ scattering in the normal state of
YBCO$_7$. Recall that the polaronic effect that arises from the
$E_I$ based Fermi-Dirac statistics (iFDS) has been successful to
explain and predict the evolution of resistivity with doping and
to enumerate the minimum valence state of multivalent dopants in
HTSC, ferromagnets and recently in doped-ferroelectrics. But iFDS
does not reveal the inadequacy of the free $e$-$ph$ scattering
directly (only indirectly). Basically, according to the $e$-$ph$
scattering, the electrons from Ba$^{2+}$ and Sr$^{2+}$ as in
Y(Ba$_{1-x}$Sr$_x$)$_2$Cu$_3$O$_7$ has the same effect on
transport measurements while iFDS points out that the kinetic
energy (KE) of the electrons from Ba$^{2+}$ is not equal with the
KE of the electrons from Sr$^{2+}$, which gives rise to
significant changes of resistivity with small doping. Again, if
one assumes KE (Ba$^{2+}$) = KE (Sr$^{2+}$), then the theory of
the $e$-$ph$ scattering is indeed applicable due to isotropy in KE
(all the free electrons have an identical KE, which eventually
defines the Fermi surfaces). Hence, to further evaluate the
incompatibility of the $e$-$ph$ scattering in YBCO$_7$, the BG
formula~\cite{tu43} stated in Eq.~\ref{eq:7} is employed to plot
the $T$-dependence of $\rho(T)$ (assuming $\tau_{e-ph}(3D)$
$\propto$ $\tau_{e-ph}(2D)$) and $\mathcal{L}(T)$.

\begin{eqnarray}
\rho_{BG} &=& \lambda_{tr} \frac{128 \pi m^*
k_BT^5}{ne^2\Theta_D^4}\int\limits_0^{\Theta_D/2T} \frac
{x^5}{\sinh^2x} dx. \label{eq:7}
\end{eqnarray}

$\lambda_{tr}$ $=$ electron-phonon coupling constant, $m^*$ =
average effective mass of the occupied carrier states, $\Theta_D$
$=$ Debye temperature, $n$ $=$ free electrons concentration. The
$\mathcal{L}(T)$ can be simply written as

\begin{eqnarray}
\mathcal{L}_{BG} &=& \kappa \lambda_{tr} \frac{128 \pi m^*
k_BT^4}{ne^2\Theta_D^4}\int\limits_0^{\Theta_D/2T} \frac
{x^5}{\sinh^2x} dx. \label{eq:8}
\end{eqnarray}

\subsection{7. Analysis}

Figure~\ref{fig:2} a) and b) depict the $T$-dependence of
$\rho(T)$ (Eq.~(\ref{eq:7})) and $\mathcal{L}(T)$ respectively.
The $\mathcal{L}(T)$ based on BG's approach after incorporating
the experimental $\kappa$ are indicated with $\vartriangle$
($\Theta_D$ = 200 K), $\circ$ ($\Theta_D$ = 300 K), and $\square$
($\Theta_D$ = 350 K). On the other hand, the experimental and iFDS
based theoretical plots (Eqs.~(\ref{eq:3}),~(\ref{eq:4})
and~(\ref{eq:6})) are shown with $\bullet$ and a solid line,
respectively in Fig.~\ref{fig:2} b). Note that in
Eq.~(\ref{eq:3}), $\alpha$ = 3 is used complying with the earlier
assumption of $T^{\alpha > 2}$. This value is reasonable since
$\alpha$ in the free $e$-$ph$ scattering of conventional metals
are known to vary between 3 and 5, depending on $T$'s range that
can be verified from Eq.~(\ref{eq:7}). The experimental
$\mathcal{L}(T)$ is obtained from the resistivity~\cite{leridon45}
and heat conductivity~\cite{sutherland42} measurements of
optimally doped YBCO.

The inverse proportionality of the theoretically determined
$\kappa$ with $T$ from Eq.~(\ref{eq:3}) is understandable since
the electrical conductivity is proportional to $1/T$ and there are
phonon contribution as well. As a result of this, $\mathcal{L}(T)$
is also inversely proportional to $T$. It is not possible to
evaluate Eq.~(\ref{eq:3}) quantitatively due to the unknown
magnitudes of $E_I$, $E_F$, $m^*_{ph}$ and $A_{e-ph}$. However,
the measured $\kappa$ in the normal state of YBCO hardly shows
strong $T$ dependence~\cite{sutherland42} indicating the existence
of some not-yet-known physical phenomena, which complicates our
understanding of HTSC generally. Anyhow, by using the
experimentally determined $\kappa$, one can verify the accuracy of
the resistivity equations (between Eq.~(\ref{eq:4})
and~(\ref{eq:7})). The former equation is entirely based on
$e$-$e$ scattering while the latter contains the essential
$e$-$ph$ scattering mechanism. To this end, the lorentz ratio
based on iFDS (Eq.~(\ref{eq:6})) and BG (Eq.~(\ref{eq:8})) are
computed using the almost $T$ independent or experimental
$\kappa$.

The iFDS model reproduces the $T$ dependence trend, remarkably
identical with the experimental data as opposed to the BG's
approach. Both iFDS and BG models with the experimental $\kappa$
have been plotted in Fig.~\ref{fig:2} b), in which the latter
model is plotted at different $\Theta_D$. Eventually, one can
convincingly state that $e$-$ph$ scattering mechanism is
significantly negligible in the electrical resistance
measurements. The plot that corresponds to Eq.~(\ref{eq:6}) with
experimental $\kappa$ is obtained using $E_I + E_F$ = 10 K, (which
is less than $T_c$ as a result of optimal or over doping) and
experimental $\kappa$ that eventually give $A$ = 1 $\times$
10$^{-8}$ $\Omega \cdot$cm. This magnitude is remarkably identical
with the optimally doped crystalline YBCO sample of Hagen {\it et
al}.~\cite{hagen22} and Leridon {\it et al}.~\cite{leridon45} that
have been calculated ($A_{Hagen,Lerridon}$ = (1.1,1.4) $\times$
10$^{-8}$ $\Omega \cdot$cm) and reported in the
Refs.~\cite{andrew5,andrew2}. Importantly, even though
Eq.~(\ref{eq:7}) can be shown to capture the experimental
$T$-linear property of $\rho(T)$, but it also fails to explain the
$T_{cr}$ above $T_c$ for slightly under doped HTSC. $T_{cr}$ is
the $T$ where $\rho(T)$ deviates upward exponentially, which has
been well explained~\cite{andrew4,andrew5,andrew6} via $E_I$ in
Eq.~(\ref{eq:4}).

\subsection{8. Conclusions}

In conclusion, iFDS based electrical resistivity (with $e$-$e$
scattering rate only) and heat conductivity (with both $e$-$e$ and
$e$-$ph$ scattering rate) models have been utilized to tackle the
$T$ dependence of Lorentz ratio in optimally doped
YBa$_2$Cu$_3$O$_7$. The computed $\mathcal{L}(T)$ with
experimental $\kappa$ overwhelmingly suggests that
Bloch-Gr\"{u}neisen formula or the inclusion of $e$-$ph$
scattering in the electrical resistivity is not suitable, at least
for YBa$_2$Cu$_3$O$_7$. On the other hand, $e$-$ph$ scattering
contributes significantly in heat conductivity that eventually
gives a reasonably acceptable picture for the experimental heat
conductivity and Lorentz ratio. Additionally, the spin Pseudogap
phenomenon have been omitted throughout so as to avoid its
inconclusive interpretations. Apart from that, the magnitudes of
the $T$-independent scattering rate constant, effective mass and
the charge carriers density are all in the acceptable range,
complying with other optimally doped YBCO single crystals as
computed previously.

\subsection*{Acknowledgments}

The author is grateful to Arulsamy Innasimuthu, Sebastiammal
Innasimuthu, Arokia Das Anthony and Cecily Arokiam of CMG-A for
their hospitality. ADA also thanks M. Sutherland for communicating
the experimental data points on heat conductivity of YBCO.

\begin{figure}
\caption {Experimental $R_H^{(ab)}(T)$ and $\rho_{ab}(T)$ (inset)
data points for YBa$_2$Cu$_3$O$_{7-\delta}$ single crystal (A1)
have been fitted using Eqs.~(\ref{eq:5}) and~(\ref{eq:4})
respectively. The former equation is computed with two $m^*$s
namely, 50$m_0$ and 73$m_0$ while the resistivity is calculated
with $A$ = 7.3 $\times$ 10$^{-7}$ $\Omega \cdot$cm.} \label{fig:1}
\end{figure}

\begin{figure}
\caption {a) Shows the BG resistivity, $\rho(T)$ plots above 90 K
for $\Theta_D$ = 350, 300 and 200 K.  Whereas, b) depicts the
theoretical plots for the BG Lorentz ratio, $\mathcal{L}_{BG}$
above 90 K with experimental heat conductivity ($\kappa$) using
Eq.~(\ref{eq:8}) with the Debye $T$, $\Theta_D$ = 350, 300 and 200
K. The calculated $\mathcal{L}(T)$ with Eq.~(\ref{eq:6}) using
experimental $\kappa$ is also plotted with $\diamond$ in b). The
theoretical solid line in b) satisfies iFDS based models namely,
Eqs.~(\ref{eq:3}),~(\ref{eq:4}) and~(\ref{eq:6}) with $\alpha$ =
3. The experimental plots indicated with $\bullet$ is obtained
from the data combined from Leridon {\it et al}.~\cite{leridon45}
and Sutherland {\it et al}.~\cite{sutherland42}.}\label{fig:2}
\end{figure}

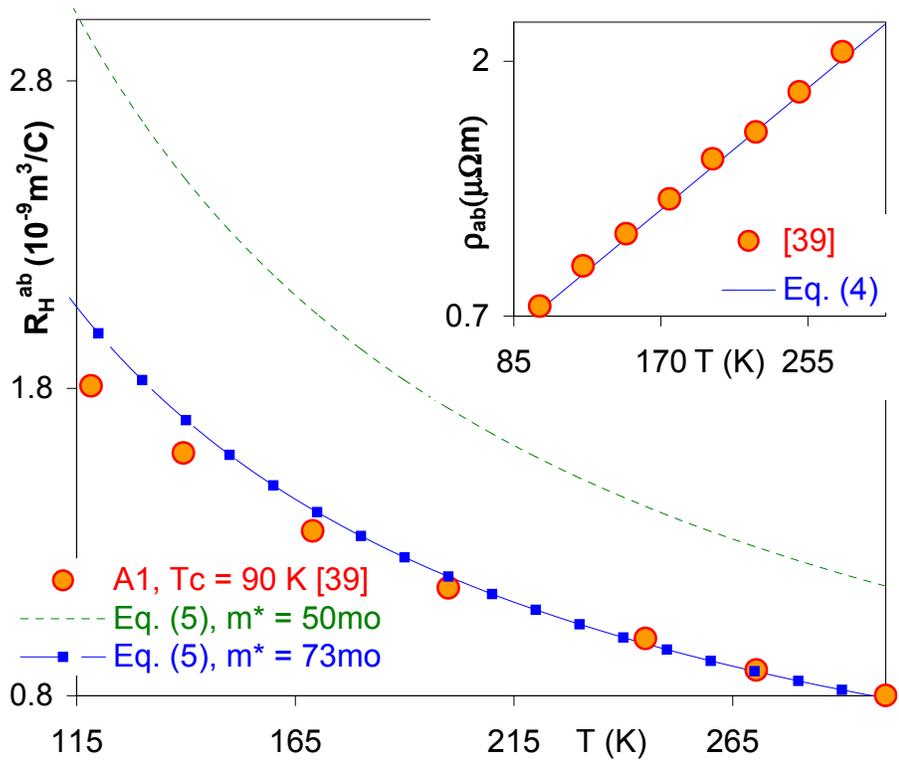
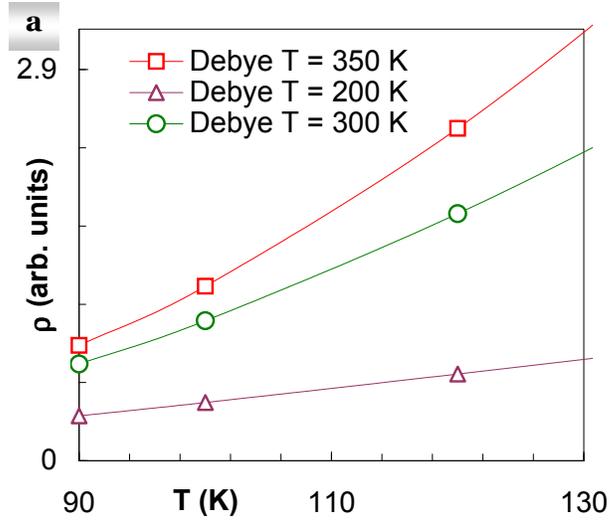
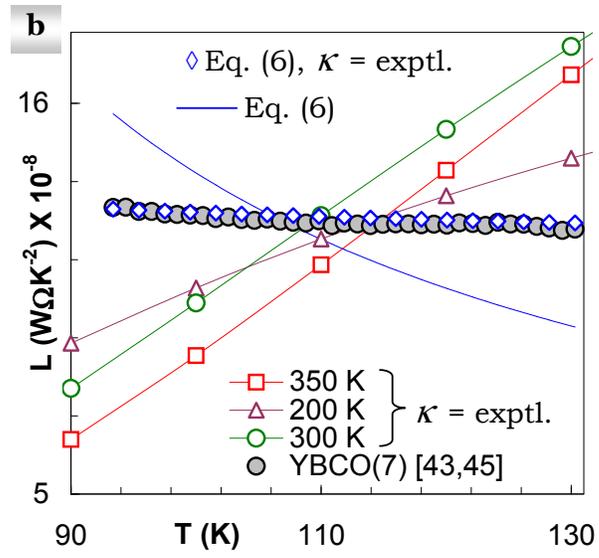
\end{document}